\documentclass [aps,prb,reprint,showpacs]{revtex4-1}
\usepackage{graphicx}
\usepackage{gensymb}
\usepackage{amsmath}
\usepackage{bm}

\usepackage{graphicx}
\usepackage{dcolumn}
\usepackage{bm}
\usepackage{txfonts}

\begin{document}

\preprint{APS/123-QED}

\title{High-Q photonic crystal cavities in all-semiconductor photonic-crystal heterostructures}

\author{Z. L. Bushell}
\author{M. Florescu}
\email{m.florescu@surrey.ac.uk}
\author{S. J. Sweeney}
\affiliation{ Advanced Technology Institute and Department of Physics, University of Surrey, Guildford, Surrey, GU2 7XH, UK}

\date{\today}

\begin{abstract}
Photonic crystal cavities enable the realization of high Q-factor and low mode-volume resonators, with typical architectures consisting of a thin suspended periodically-patterned layer to maximize confinement of light by strong index guiding. We investigate a heterostructure-based approach comprising a high refractive index core and lower refractive index cladding layers. Whilst confinement typically decreases with decreasing index contrast between the core and cladding layers, we show that, counter-intuitively, due to the confinement provided by the photonic band structure in the cladding layers, it becomes possible to achieve Q-factors $>10^4$ with only a small refractive index contrast. This opens up new opportunities for implementing high Q-factor cavities in conventional semiconductor heterostructures, with direct applications to the design of electrically-pumped nano-cavity lasers using conventional fabrication approaches.

\begin{description}
\item[PACS numbers]
42.70.Qs,
41.20.Jb,
42.55.Tv,
78.67.Pt
\end{description}
\end{abstract}

\maketitle


The development of low-threshold, compact laser sources is highly relevant for the implementation of photonic integrated circuits based on III-V or silicon platforms (or hybrid combinations), with many applications including high-bandwidth communications, sensing and imaging \cite{Chaisakul2014,Smit2012,Kita2008,Abe2015}. Photonic crystal (PhC) nanocavities are ideal candidates for such devices due to their very small mode volume, high Q-factor and strong light-matter interaction \cite{Foresi1997}.

A number of optically pumped PhC cavity lasers have been demonstrated, mostly using a two-dimensional lattice of air holes perforating a thin (half-wavelength) semiconductor slab surrounded above and below by air \cite{Painter1999,Matsuo2010,Nomura2008}. The PhC lattice provides light confinement within the plane of the material, whilst the air-clad membrane design minimizes the out of the plane losses by providing strong index guiding. However, there are several major drawbacks of this approach including poor thermal conductivity, which has led to difficulties in achieving continuous-wave (cw) operation at room temperature, as well as the high manufacturing cost and complexity. There are some examples of PhC cavities bonded to low refractive index substrate layers, such as sapphire or silicon dioxide, which show decreased thermal resistance \cite{Shih2007,Hwang2000} but these are electrically insulating. The use of either air-clad membranes and/or insulating substrate layers causes significant drawbacks in adapting to an electrically pumped design. Recent examples have used ion-implantation to create a lateral p-i-n junction \cite{Ellis2011,Takeda2013} which, although successful, make fabrication considerably more complex than a standard vertical junction that can be grown epitaxially.

One approach to overcome these challenges is to incorporate additional cladding layers above and below the active region, rather than using a free-standing thin membrane. These cladding layers aid thermal conductivity, structural robustness, and, for doped semiconducting layers, also allow for electrically pumped designs without the need for ion-implantation. There has been very little previous work in this area \cite{Johnson1999,Benisty2000,Mock2010}; Ref. \cite{Mock2010} considered a particular photonic crystal cavity design and a few combinations of cladding layers but did not explore a relevant range of refractive index combinations actually possible in realistic material systems, most likely due to the anticipated monotonic variation of the quality factor with refractive index contrast. 

\begin{figure}
\includegraphics[width=\linewidth]{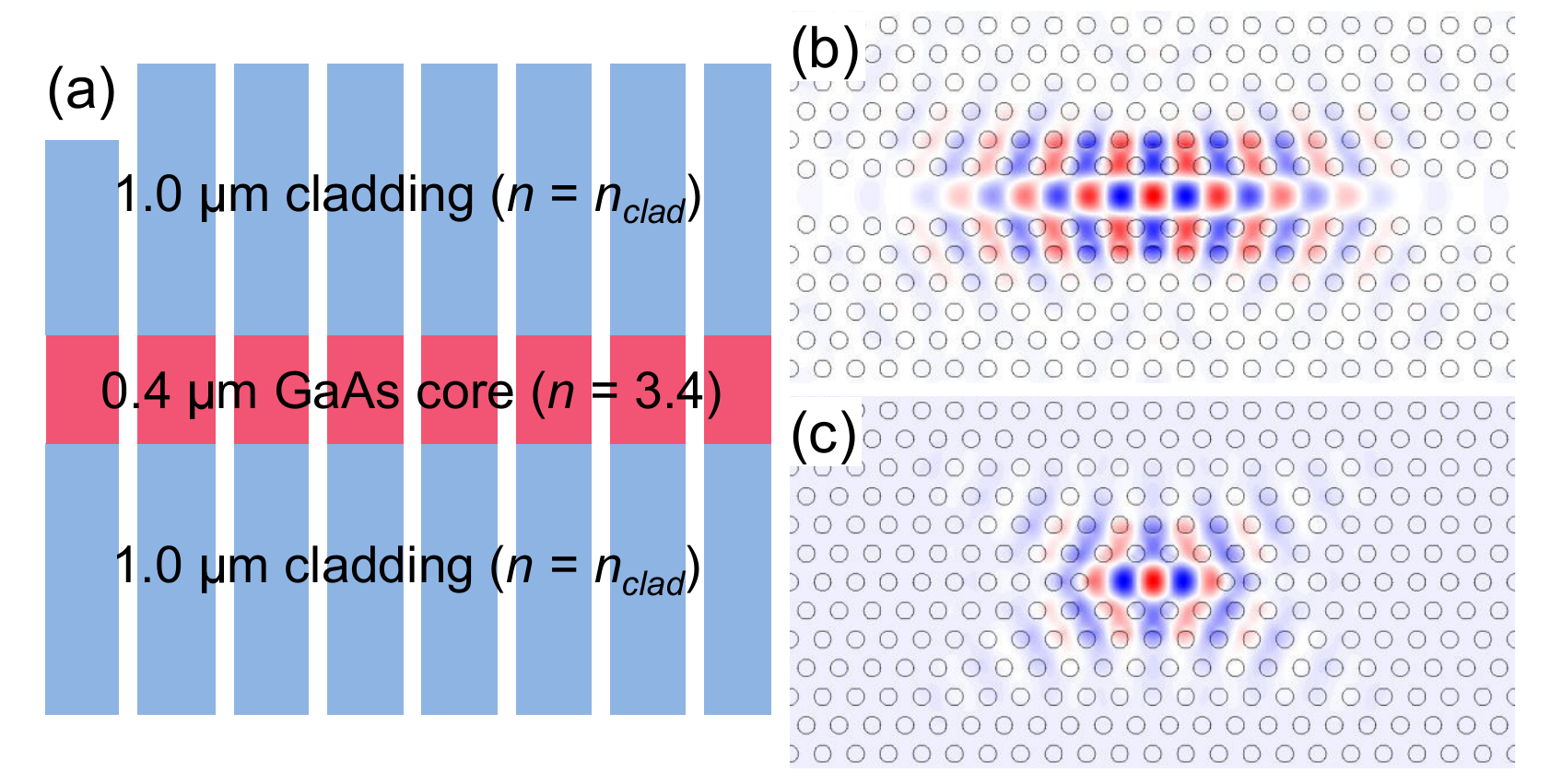}
\includegraphics[width=\linewidth]{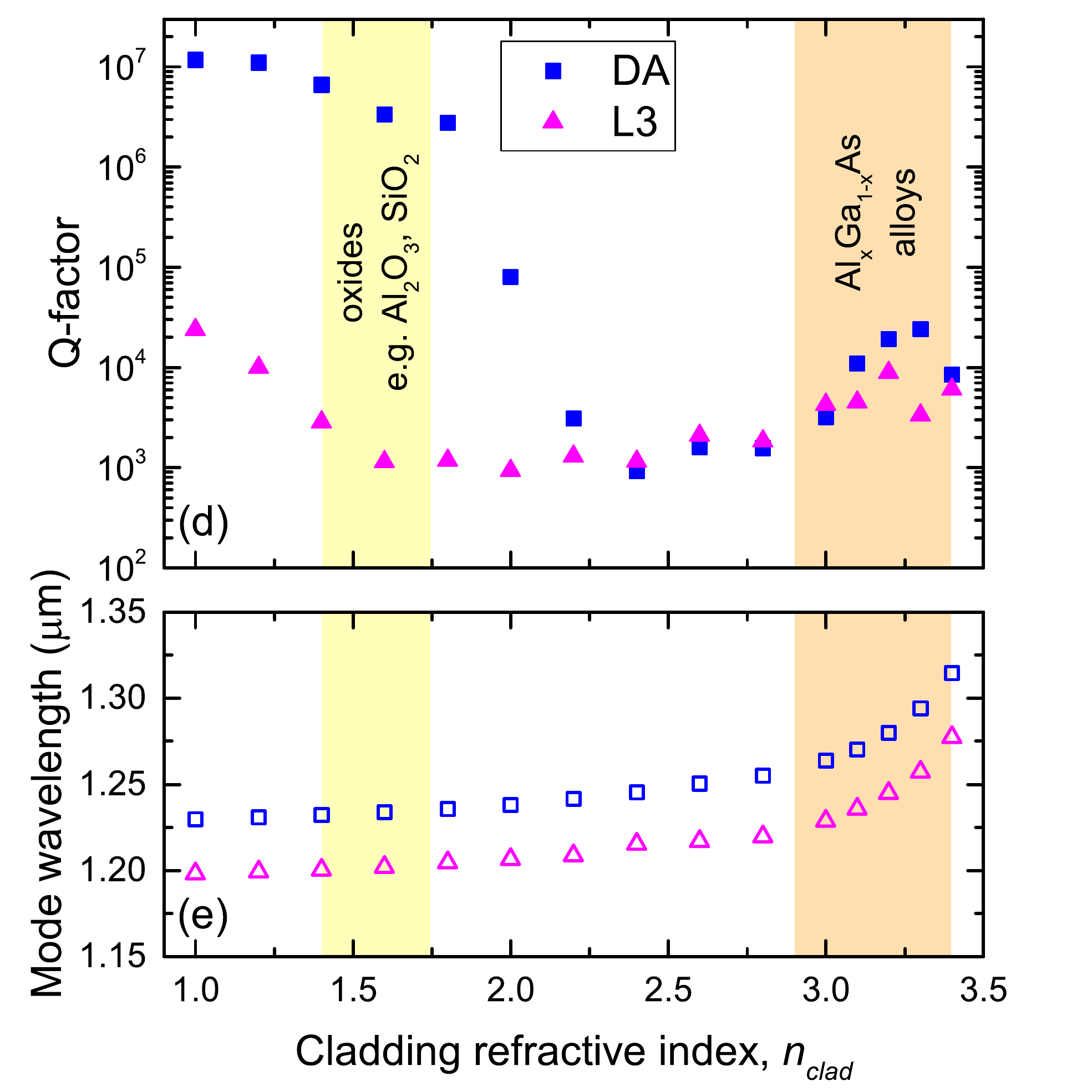}
\caption{ \label{fig:fig1} (a) Vertical cross-section of the slab layer architecture. (b),(c) Electric-field profiles of fundamental cavity mode in DA and L3 cavities, respectively. (d) Q-factor and (e) fundamental mode wavelength of DA (squares) and L3 (triangles) photonic crystal nanocavities in 400 nm GaAs ($n = 3.4$) slab surrounded by 1 $\muup$μm thick cladding layers as function of cladding refractive index.}
\end{figure}

In this article, we explore different combinations of nanocavity designs and material layers using 3-dimensional finite difference time domain (FDTD) simulations. We consider a multilayer slab structure with the air holes of the PhC lattice extended completely through the core and cladding layers, as illustrated in Fig.  \ref{fig:fig1}(a). The core layer is 400 nm thick and has a fixed refractive index of 3.4, which corresponds to that of GaAs in the near-infrared, whilst the refractive index $n_{clad}$ of the 1 $\muup$m thick cladding layers is varied from 1.0 -- 3.4. Such structures require high aspect ratio etching, as has already been demonstrated in the literaure \cite{Avary2002,Kim2012b,Kita2011,Moosburger2002}. We consider two well-established high Q-factor designs: the modified L3 \cite{Kuramochi2014} and dispersion adapted (DA) \cite{Welna2013} cavities. The DA cavity is of particular interest here as it is better equipped to minimize out-of-plane scattering losses and hence it is less affected by reduced total internal reflection at the upper and lower interfaces. The L3 design is a conventional photonic crystal cavity architecture and is chosen as a baseline for comparison.

The photonic crystal consists of a triangular lattice with period $a = 270$ nm and radius $r = 0.25a$. The lattice covers an area of $50a$ x $30a$, and the slab material has a finite area of $52a$ x $30a$ with the simulation region boundaries lying outside the slab. The DA cavity is formed from a W1 waveguide, where one complete row of holes is removed, and pairs of holes along the edge of the waveguide are shifted by the distances given for the $h_020$ cavity described in Ref. \cite{welnaphd}. The L3 cavity has 3 missing holes and those at each end of the cavity are shifted outwards by $dx = 0.22a$ \cite{Kuramochi2014}. The cavity designs and electric-field profiles of the fundamental resonant mode in each case are shown in Fig.  \ref{fig:fig1}(b) and (c) for the DA and L3 cavities, respectively.

FDTD simulations of the structures were performed with PML boundary conditions in all three directions \cite{Kim2012b,Kim2012}, assuming 7$\mu$m of air between photonic-crystal slab and boundary. The resonant modes were identified from the peaks in the frequency spectrum $E(\omega)$, whilst the Q-factor was derived from the corresponding exponential time decay of the electric field. Figure \ref{fig:fig1}(d) shows the dependence of the cavity Q-factor on the cladding refractive index. The same fundamental cavity modes, shown in Fig. \ref{fig:fig1}(b) and (c), are tracked throughout and plotted in Fig. \ref{fig:fig1}(e). There is an expected gradual shift to longer wavelengths as the average refractive index of the system increases. For the air-clad ($n_{clad} = 1.0$) case, the DA cavity has a very high Q-factor of $\sim$1x$10^7$, as expected given that this was designed to maximize Q in the 400 nm slab \cite{welnaphd}. The L3 cavity shows significantly worse performance with a Q-factor of only $\sim$2x$10^4$, which is again expected as no attempt has been made to optimize the design for the chosen material dimensions. As the refractive index of the cladding layers is increased from its air value, the Q-factor of both cavity designs initially decreases, in agreement with the decreased vertical confinement of the mode due to reduced total internal reflection at the core-cladding interfaces. The DA cavity still performs well, with Q $> 10^6$, when $n_{clad}$ is within the range 1.4 -- 1.7 corresponding to typical oxide materials such as Al$_2$O$_3$. The use of oxide layers is known to improve thermal conductivity and structural robustness compared to an air-clad membrane but does not address the challenges related to electrical injection, since they are highly electrically resistive. 

As the cladding refractive index is increased further, above 2.4, there is an increase in Q-factor for both cavity designs. This reaches a peak value $>10^4$ for the DA cavity when $n_{clad}$ is in the range 3.2 -- 3.3. The L3 cavity Q-factor peaks at $\sim$9x$10^3$ when $n_{clad} = 3.2$. These are a remarkable results since these values lie within the highlighted range of $n$ for AlGaAs alloys, opening up the possibility to implement high-Q photonic crystal cavities in standard epitaxial semiconductor heterostructures in GaAs/AlGaAs material systems. We note, however, that this behavior seems counterintuitive, as one would expect the Q-factor to monotonically degrade as index guiding is reduced and a greater amount of light is able to leak into the cladding.

\begin{figure}
\includegraphics[width=\linewidth]{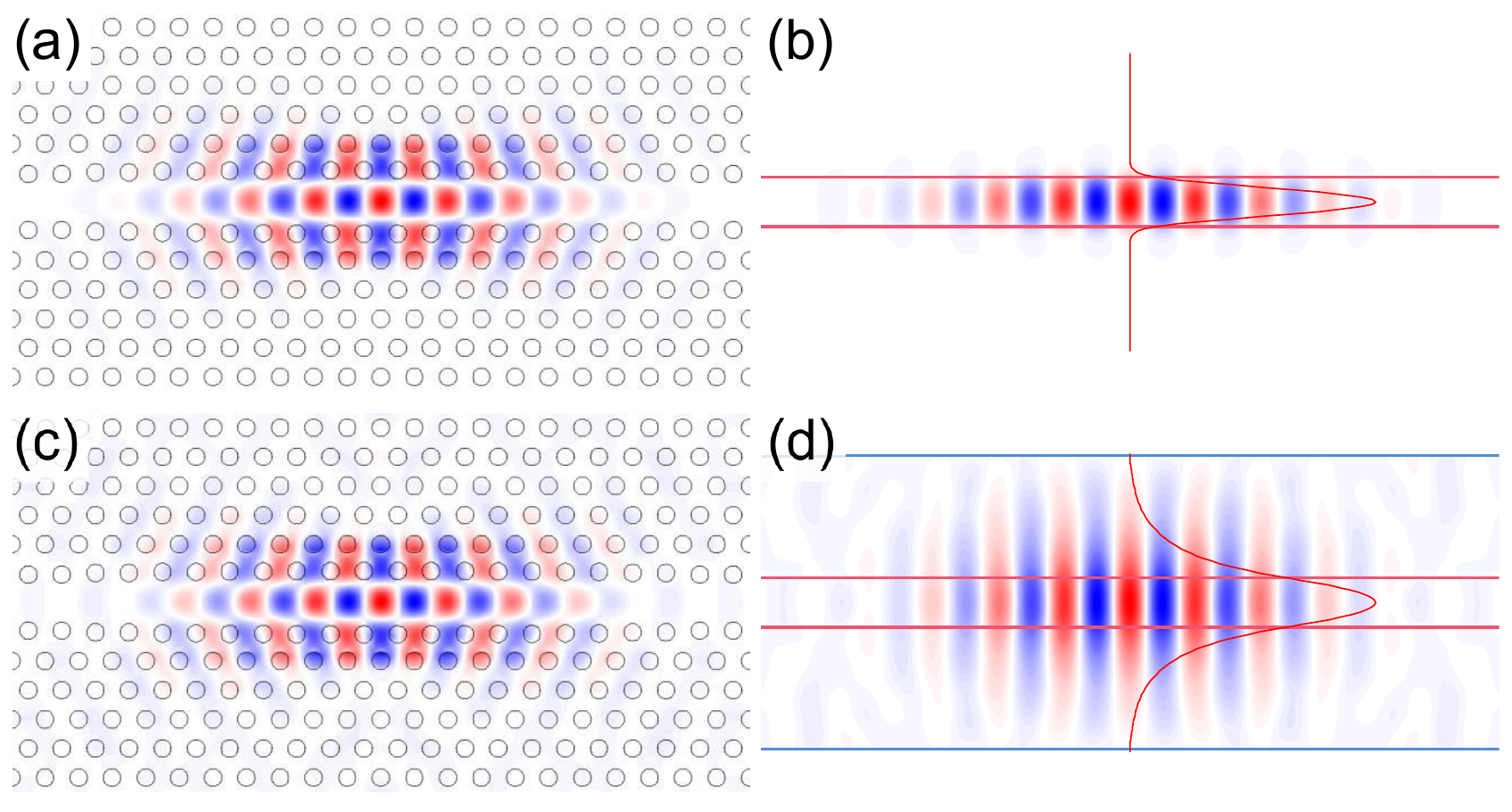}
\caption{ \label{fig:extra_DA} Electric-field profiles in XY plane (a,c) and XZ plane, overlaid with line profile through cavity center (b,d) for DA cavity with $n_{clad} = 1.0$ (a,b) and 3.3 (c,d).}
\end{figure}

To explore the origin of this unexpected behavior we begin by examining the electric-field profiles to establish how the spatial distribution of the mode changes with the refractive index of the cladding. Cross sections showing the electric-field profiles of the DA cavity in both horizontal and vertical planes through the center of the cavity are shown in Fig. \ref{fig:extra_DA}. The horizontal cross sections in Fig. \ref{fig:extra_DA}(a),(c) illustrate that the mode remains well confined to the cavity region, with minimal change to the mode profile as the cladding index increases. The vertical cross sections in Fig. \ref{fig:extra_DA}(b),(d) show a greater spread of the mode into the cladding as the refractive index increases due to the reduced index guiding. The overlaid line profile through the cavity center and its optical confinement factor are typical of GaAs/AlGaAs waveguides \cite{Wu2010,Givens1992}, indicating that good overlap of the optical mode with the active region is maintained.

\begin{figure}
\includegraphics[width=\linewidth]{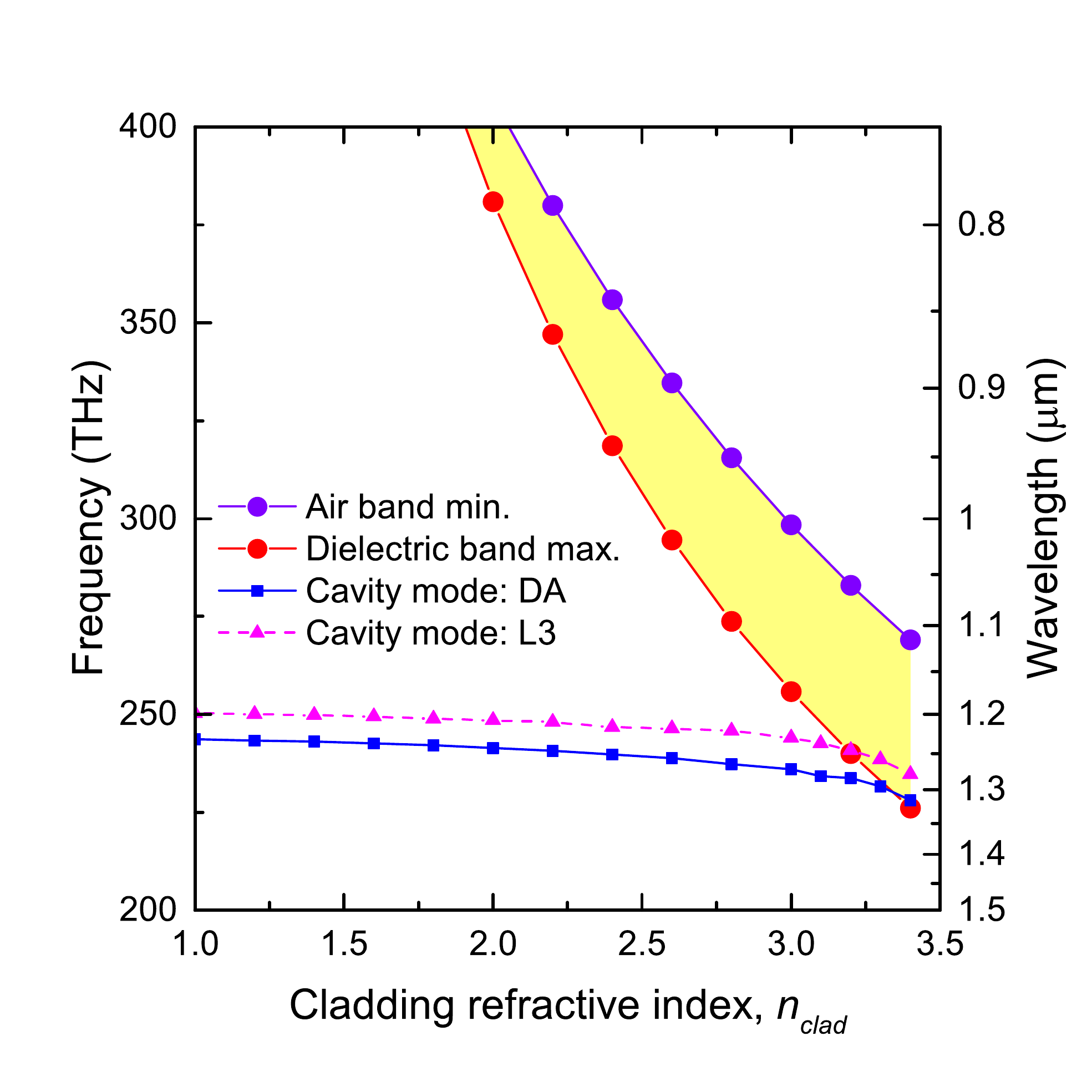}
\caption{\label{fig:fig2} Band edges (red \& purple circles) and band gap (shaded yellow) of 2D photonic crystal as function of material refractive index. Resonant mode frequencies for DA (squares) and L3 (triangles) cavities in full multilayer slab structures shown for comparison.}

\end{figure}

Since the peak Q-factor occurs for a small difference between the refractive indices of the core and cladding, we suggest that the photonic band gaps of the core and cladding layers may overlap in frequency, such that the photonic band gap of the cladding prevents propagation of the resonant mode and consequent degradation of the Q-factor. As an initial test of this hypothesis, the band structure of a 2D PhC with the same lattice parameters as used in the full structure was computed for each refractive index value. Figure \ref{fig:fig2} shows the resulting PBG frequency range along with the resonant cavity mode frequencies from the FDTD simulations. In the region of $n \geq3.2$, the resonant mode of the L3 cavity lies within the 2D band gap of the cladding layers, whilst the DA cavity resonance is just below the band edge of the dielectric band. This shows that the mode confinement is aided by the cladding PhC properties. In the case where the cavity mode lies within the PBG of the cladding, the mode is unable to propagate away in-plane through the cladding. When it lies just below the dielectric band edge, the dispersion curve is relatively flat in this frequency range so the group velocity of the propagating mode tends to zero, and as such the cavity mode remains localized to the cavity region \cite{florescu2010}. This confining mechanism explains how a high-Q mode is maintained by reducing propagation of the light away in-plane, even though it is able to spread vertically into the cladding due to the low refractive index contrast.

This analysis treats the cladding as an infinite 2D slab, which is not the case within the actual system. The cladding layers have a finite thickness and are also in an asymmetric arrangement due to the presence of the core layer at one interface and air at the other. It is known that there is still a pseudo-gap effect in optically thick PhC slabs, where cavity modes have only minimal coupling to the extended Bloch modes, that can lead to high-Q resonant states \cite{Kim2012}. This effect can be further enhanced in slabs with reflective (e.g. metallic or semiconductor/air) boundaries in the horizontal plane, due to alterations of the density of Bloch modes in momentum space \cite{Kim2012}. The structures investigated here fulfil both of these criteria, being both optically thick and having a finite horizontal extent with semiconductor/air interfaces within the simulation region. It is therefore expected that a similar effect occurs in this case and the argument of minimal propagation through the cladding would hold. To demonstrate this we undertook further analysis of the FDTD simulation results to investigate this hypothesis.

\begin{figure}
\includegraphics[width=\linewidth]{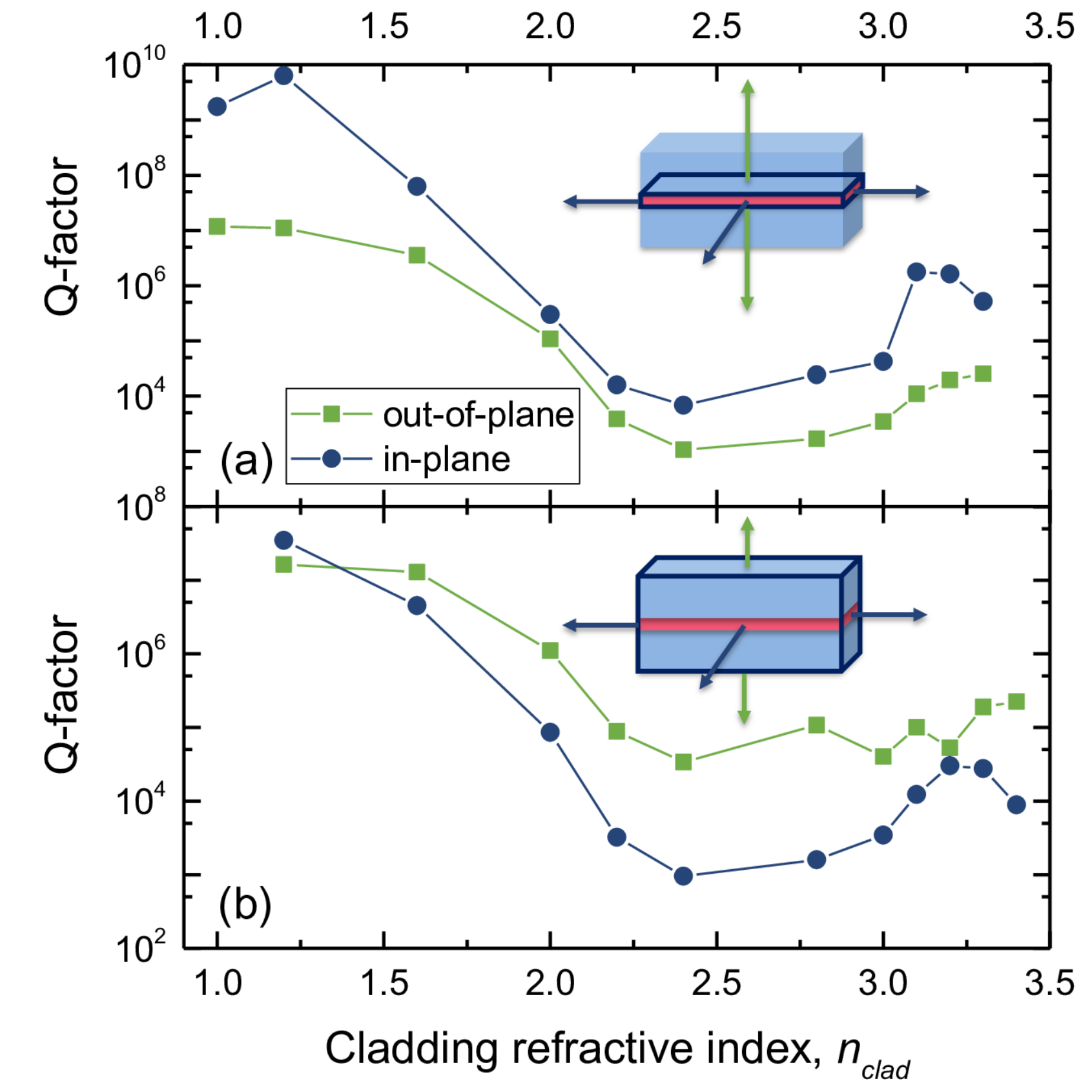}
\caption{ \label{fig:fig3} Directional Q-factors of DA cavity, showing contributions from losses through surfaces of (a) core layer only and (b) whole multilayer structure, as illustrated in inset.}
\end{figure}

The relative contributions to losses from the DA cavity are determined by monitoring the power flow through various surfaces surrounding the cavity. The power flow is obtained from the FDTD simulation by integrating the time-averaged Poynting vector across the surface of interest. This is converted to the directional Q-factor as described in Ref. \cite{Mock2010}. In Fig. \ref{fig:fig3}(a) we first consider the loss from the core layer only and break it down into two components: the out-of-plane loss vertically into the cladding layers and the in-plane loss through the photonic crystal lattice. The lowest Q-factor and therefore dominant contribution to loss is in the out-of-plane direction and increases as $n_{clad}$ increases due to the decreased index guiding. There is a small increase in this out-of-plane Q-factor in the region of $n_{clad} \geq 3.2$  where the peak Q-factor occurs. 

By contrast, in Fig. \ref{fig:fig3}(b) we consider the loss through the outer surfaces of the entire structure. We see here that once light reaches the cladding, the lowest Q-factor and therefore main loss is actually the in-plane loss, rather than that through the upper and lower surfaces. There is a strong increase of more than one order of magnitude in this component of the Q-factor going towards the $n_{clad} = 3.2 - 3.3$ region where the overall Q-factor peaks, showing the propagation of light through the cladding has reduced. This further supports our argument that photonic band structure effects in the cladding are reducing propagation of the resonant mode and are thereby increasing the Q-factor.

\begin{figure}
\includegraphics[width=\linewidth]{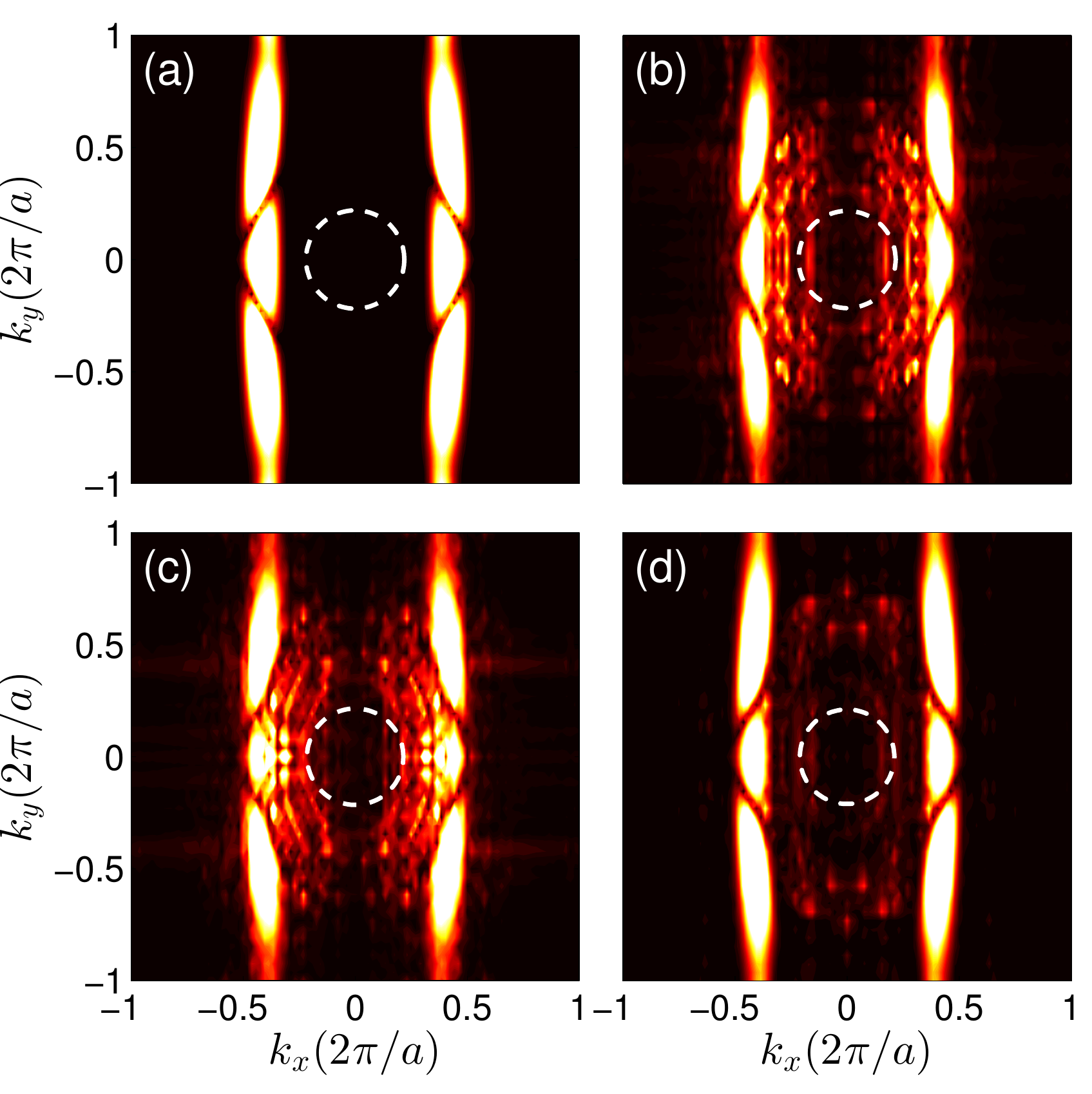}
\caption{ \label{fig:fig4} Spectra of the Fourier components of the electric-field distribution for the DA cavity mode within the cladding layer, with $n_{clad}$  = 1.2 (a), 2.4 (b), 2.8 (c) and 3.2 (d). Light circle $k = \omega c$ indicated with dashed white line.}
\end{figure}

\begin{table}
\caption{\label{tab:tab1} Fraction of Fourier transform (FT) components falling within light cone for spectra shown in Fig. \ref{fig:fig4}.}

\begin{ruledtabular}
\begin{tabular}{lcccr}
$n_{clad}$ & $Q_{in-plane}$ & $Q_{out-of-plane}$ & FT in light cone (\%) \\ \hline\noalign{\smallskip}
1.2 & 3.5x$10^7$ & 1.6x$10^7$ & 0.04\\ 
2.4 & 9.5x$10^2$ & 3.4x$10^4$ & 0.49\\  
2.8 & 1.6x$10^3$ & 1.1x$10^4$ & 0.38\\ 
3.2 & 3.0x$10^4$ & 5.3x$10^4$ & 0.22\\
\end{tabular}
\end{ruledtabular}
\end{table}

We also analyzed the spatially Fourier transformed in-plane electric-fields. Only Fourier components lying with the light circle defined by $k$$<$$\omega c$ can radiate into the far field, so comparing the Fourier spectra can give information about loss from the modes \cite{amoah2015}. Additionally, if components overlap in momentum space with the extended Bloch modes of the cladding, the cavity mode is able to couple to those extended modes and propagate away. As it is the behavior of the mode once it spreads to the cladding that is of interest, Fourier transforms are taken of the electric-field profile within the cladding, close to the core-cladding interfaces. Examples of the Fourier spectra for different values of $n_{clad}$ are shown in Figs. \ref{fig:fig4}(a)-(d). It can be seen that in the high-Q cases of $n_{clad} = 1.2$ and 3.2, given in Fig. \ref{fig:fig4}(a) and (d) respectively, there are fewer Fourier components within the light circle and also in the immediate surrounding region. In contrast, the large number of high-intensity points present in Fig. \ref{fig:fig4}(b) and (c) indicate the significant coupling of the cavity mode to the extended Bloch modes in the cladding. This coupling enables the escape of the localized mode away from the cavity and leads to reduced Q-factors. In order to make a quantitative analysis, all Fourier spectra were normalized and the fraction of the components falling within the light circle was then calculated by integrating the intensity within this area. The results presented in Table. \ref{tab:tab1} show that the fraction of Fourier components within the light cone is reduced by around a factor of two in the high Q-factor $n_{clad} = 3.2$ case compared to the $n_{clad} = 2.4$ and 2.8 cases with low Q-factor. This clearly demonstrates the qualitative trend shown by the profiles of Fig. \ref{fig:fig4}. This argument again validates the hypothesis that the high Q-factor in the $n_{clad} = 3.2 - 3.3$ range is caused by photonic band structure effects in the cladding, with minimal coupling of the cavity mode to extended modes reducing propagation of light away from the cavity.

In summary, our study demonstrates that it is possible to maintain Q-factors $\sim10^6$ with cladding layers corresponding to typical oxide materials and, more importantly, $>10^4$ with cladding layers corresponding to typical semiconductors such as AlGaAs alloys, each surrounding a GaAs ($n$ = 3.4) core layer. The relatively high Q-factor in the low index contrast  $n_{clad}= 3.2 - 3.3$ cases is explained in terms of band structure and pseudo-PBG effects in the cladding layers. These effects prevent propagation of the light in-plane through the cladding, despite the fact that the index guiding responsible for the vertical confinement of the light to the core layer is reduced. This counterintuitive behavior can be exploited to design PhC nanocavities in low refractive index contrast multilayer slabs, for example in standard GaAs/AlGaAs semiconductor heterostructures. This opens up a new design possibility for nanocavity lasers with improved properties and ease of manufacture compared to existing air-clad membrane designs.

\begin{acknowledgements}
Z.L.B. gratefully acknowledges support from the Marion Redfearn Scholarship and Advanced Technology Institute Scholarship. We also gratefully acknowledge funding under EPSRC (United Kingdom) grants EP/H005587/1, EP/L02263X/1 (EP/M008576/1) and EP/M027791/1. The authors confirm that data underlying the findings are available without restriction. Details of the data and how to request access are available from the University of Surrey publications repository: http://epubs.surrey.ac.uk/809461/.
\end{acknowledgements}

\providecommand{\noopsort}[1]{}\providecommand{\singleletter}[1]{#1}%

\end{document}